\documentclass[epjST]{svjour}

\usepackage{graphics}
\usepackage{epstopdf}
\usepackage{color}
\usepackage{textcomp}

\begin{document}

\title{Visualizing driving forces of spatially extended systems using the recurrence plot framework}
\subtitle{}
\author{M. Riedl\inst{1}\fnmsep\thanks{\email{maik.riedl@pik-podsdam.de}} \and N. Marwan\inst{1} \and  J. Kurths\inst{1}\inst{2}}

\institute{Potsdam Institute for Climate Impact Research (PIK), Germany \and Humboldt-Universität zu Berlin, Germany}


\abstract{
The increasing availability of highly resolved spatio-temporal data leads to new opportunities as well as challenges in many scientific disciplines such as climatology, ecology or epidemiology. 
This allows more detailed insights into the investigated spatially extended systems.
However, this development needs advanced techniques of data analysis which go beyond standard linear tools since the more precise consideration often reveals nonlinear phenomena, for example threshold effects.
One of these tools is the recurrence plot approach which has been successfully applied to the description of complex systems.
Using this technique's power of visualization, we propose the 
analysis of the local minima of the underlying distance matrix in order to display driving forces of spatially extended systems.
The potential of this novel idea is demonstrated by the analysis of the chlorophyll concentration and the sea surface temperature in the Southern California Bight.
We are able not only to 
confirm the influence of El Ni\~no events on the phytoplankton growth in this region but also to 
confirm two discussed regime shifts in the California current system.
This new finding underlines the power of the proposed approach and promises new insights into other complex systems. 
} 

\maketitle

\section{Introduction}
\label{intro}
In many scientific disciplines, such as climatology, ecology, neuroscience or epidemiology, the increasing availability of highly resolved spatio-temporal data allows more detailed insights into the behavior of spatially extended systems of interest.
However, this development also leads to the challenging question for advanced analytical techniques in order to describe and to quantify the now revealing phenomena which often have a nonlinear character, e.g. threshold effects.
The unabated publishing of new analytical tools for this purpose proves this methodological progression.
In the last decades, the recurrence plot (RP) has emerged as one successful concept in the description of nonlinear phenomena and complex systems \cite{eckmann1987},\cite{marwan2007} but their application to spatio-temporal problems is only in the beginning.
In previous studies we have proposed the high dimensional RP and the mapogram-based RP (MRP), respectively, in order to bring the powerful concept to the analysis of spatially extended systems \cite{marwan2015},\cite{riedl2015}.
The RP framework enables us not only to quantify the dynamics of these systems by means of the library of recurrence quantification analysis \cite{marwan2007} and recurrence network analysis \cite{donner2011}, but also to visualize the underlying high-dimensional state space.
An example of this family of tools is the thresholded meta RP which was proposed by Casdagli \cite{casdagli1997} for visualizing driving forces of a system.
We extend the thresholded meta RP by using a kernel estimator instead of a histogram in order to take into account even fast changes of the force and by focusing on the underlying distance matrix avoiding gaps in the RP which results from regime shifts.
The potential of this novel idea is demonstrated by the application to remotely sensed data of the Southern California Bight (SCB), one of the most productive marine ecosystems of the world \cite{wang2015}, looking for hidden determinants of the complex phytoplankton's growth in this system.
In particular, the data are the chlorophyll concentration (CHL), a proxy of the phytoplankton's concentration, and the sea surface temperature (SST), a proxy of the hydrological state of the California Current System (CCS).
Influences at three different time scales are discussed for this system \cite{venrick2012}.
Influences at longer time scales are annual determinants which are dominated by the coastal wind-driven upwelling which typically occurs in Spring.
Other influences are short-term interannual influences, for example the El Ni\~no-Southern Oscillation (ENSO) where El Ni\~no events decrease the upwelling and increase the SST.
A previous study only indicates a weak effect of the El Ni\~no events on the CHL whereas La Ni\~na periods seem ineffective except the sharp events after the strong El Ni\~nos 1982$-$83 and 1997$-$98 \cite{venrick2012}.
Finally, long-term influences enclose cycles of 10 years or longer as well as irregular regime shifts, e.g. the major regime shift 1976$-$77.  


\section{Method}

\subsection{Recurrence plot}

The recurrence plot (RP) visualizes the recurrences of a state of a dynamical system.
The temporal evolution of such dynamical system is given by its trajectory $\{\vec{x}_i\}^N_{i=1}$ in the system's phase space.
Then, the corresponding RP is based on the recurrence matrix:
\begin{equation} \label{recurrence}
R_{i,j}(\epsilon)=\Theta(\epsilon-\|\vec{x}_i-\vec{x}_j\|).
\end{equation}
$i$ and $j$ are the indexes of the observed states and go from 1 to $N$, the number of observed states.
$\|\cdot\|$ denotes a norm and $\Theta$ is the Heaviside function.
In the RP, the values 1 in the recurrence matrix are displayed by black dots which show that the trajectory comes close (defined by the threshold $\epsilon$) to a previous state \cite{marwan2007}.

\begin{figure}
\resizebox{0.75\paperwidth}{!}{\includegraphics{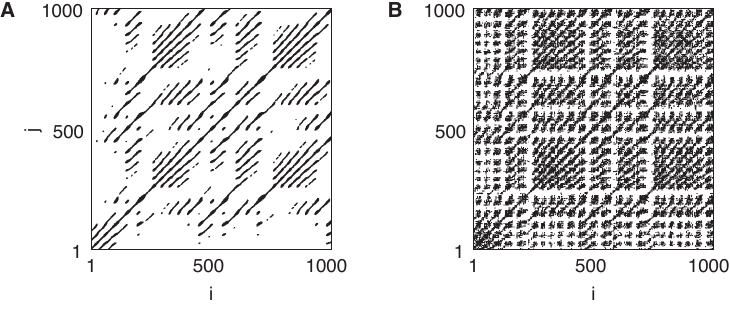}}
\caption{(A) Recurrence plot of the simulated Lorenz-system (c.f. App.~\ref{lorenz}); (B) Recurrence plot of the same simulation with additive noise (white noise added to each component; signal-to-noise ratio: 2.5). The recurrence rate is fixed to 5$\%$ in both cases.}
\label{fig2}
\end{figure}

A serious problem when constructing the RP is the contamination of the data by noise (Fig.~\ref{fig2}). 
Comparing the RP of the noisy Lorenz data with the RP of the clean Lorenz data, we find that the diagonal lines resolve into clouds of points (Fig.~\ref{fig2}).
Further, the patchy like structure with and without points vanishes, i.e. the points are more homogeneously distributed over the whole plot.
Surprisingly, the last effect reveals the hidden structure of the underlying distance matrix which allows a qualitative description of the dynamics in the sparse regions of the undisturbed RP (Fig.~\ref{fig2}A).

\subsection{Thresholded meta recurrence plot}

\begin{figure}
\resizebox{0.75\paperwidth}{!}{\includegraphics{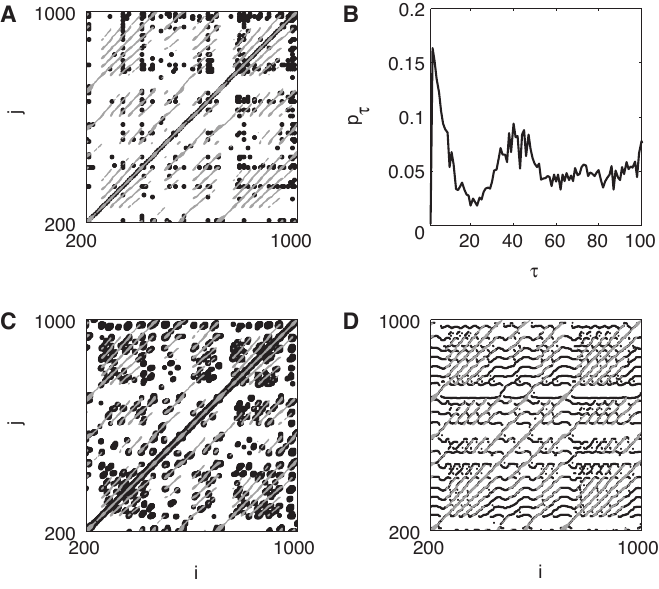}}
\caption{(A) Black points display the thresholded meta RP of the RP in Fig.~\ref{fig2}B; (B) Probability of recurrence after $\tau$ time steps, $p_{\tau}$ (Eq.~\ref{eq10}), of the RP in Fig.~\ref{fig2}B. The remarkable local minimum at 20 determines the bandwidth of the kernel used in C and D. (C) Black points display the thresholded meta RP using a kernel estimation instead of the histogram as used in A. (D) The column-wise position of the minima in the kernel smoothed distance matrix of the RP in Fig.~\ref{fig2}C. For comparison, gray points in A, C, and D show the undisturbed RP from Fig.~\ref{fig2}A.}
\label{fig3}
\end{figure}

Casdagli \cite{casdagli1997} proposed the thresholded meta RP in order to remove the noise and to reveal footprints of the system's drivers.  
The thresholded meta RP results from covering of the noisy RP by means of $N_b$ non-overlapping squares building the bins of a histogram $h_{kl}$ with $k,l=1$,...,$N_b$, the indexes of the bins.
$h_{kl}$ is the relative frequency of points in the bin. 
On the basis of this histogram, the new distance matrix is calculated by means of 
\begin{equation}\label{histEst}D^{hist}_{k,l}=(h_{k,k}+h_{l,l}-2h_{k,l})/\epsilon^2\end{equation}
which is further thresholded.
$\epsilon$ is the threshold of Eq.~\ref{recurrence} constructing the starting RP. 
The thresholded meta RP of the RP in Fig.~\ref{fig2}B is displayed in Fig.~\ref{fig3}A which only reflects the global patchy like structure of the undisturbed RP (Fig.~\ref{fig2}A).
Here, the bin width of 8 time steps is manually selected.
The recurrence rate is fixed to 5$\%$.
Although without an external driving force, the example demonstrates the effect of the thresholded meta RP to reveal long-term dynamics.   

However, the application of the thresholded meta RP is limited to slowly varying dynamics. 
One cause of this limitation is the use of histograms which does not adapt to the distribution of the recurrence points and therefore needs larger bin sizes.
So, we extend this approach by applying a kernel estimation of the point distribution instead of the histogram in order to overcome this problem.
We estimate the point distribution of the noisy recurrence plot by means of a 2D Epanechnikov kernel (Eq.~\ref{kernelEst})
\begin{equation}\label{kernelEst}d_{i,j}=\frac{1}{N^2}\sum_{i',j'=1}^{N}{K_w(\frac{i-i'}{w},\frac{j-j'}{w})R_{i,j}}\end{equation}
This kernel $K_w$ has a bell-like shape as the Gaussian kernel but has a bounded domain which avoids long-range effects \cite{silverman1986}.
The 2D Epanechnikov kernel is defined by
\begin{equation} \label{kernel}
K_w(x,y)= \left\{ \begin{array}{ll}
3(1-((x/w)^2+(y/w)^2))/4 & \mbox{if }\|(x/w)^2+(y/w)^2\|\leq 1\\
0               & \mbox{otherwise}.
\end{array} \right.
\end{equation}  
$w$ is the bandwidth of the kernel and is determined by the half of the first minimum of the probability of recurrence after $\tau$ time steps which is defined by: 
\begin{equation}\label{eq10}
p_{\tau}=\frac{1}{N-\tau}\sum_{i=1}^{N-\tau}{R_{i,i+\tau}}.
\end{equation}
$N$ is the size of the recurrence matrix $\{R_{i,j}\}_{i,j=1,...,N}$ from Eq.~\ref{recurrence} \cite{marwan2007}.
That is, $p_{\tau}$ is the relative frequency of the points in the $\tau$-th diagonal of the RP (e.g. Fig.~\ref{fig3}B).
The picked minimum of $p_{\tau}$ guarantees a high level of smoothing and the resolution of the diagonal structures in the thresholded meta RP.
From the kernel estimated distribution $d_{i,j}$ in Eq.~\ref{kernelEst}, we get the distance matrix 
\begin{equation}D^{kernel}_{i,j}=(d_{i,i}+d_{j,j}-2d_{i,j})/\epsilon^2\end{equation}
as in the case of the histogram based approach (Eq.~\ref{histEst}).
The 
extended version of the thresholded meta RP resembles the undisturbed recurrence plot much better than the original, histogram based approach.
Although the same recurrence rate, the number of the recurrence points is much higher since the down sampling by means of the histogram is not necessary anymore.

\subsection{Local minima of the distance matrix\label{newapproach}}

A second adaptation of the approach by Casdagli \cite{casdagli1997} is done in order to overcome the sparse regions of the RP which are caused by instationarities (e.g. Fig.~\ref{fig2}A).
For this purpose, we consider the local minima in the columns of the underlying distance matrix ($D_{i-1,j}>D_{i,j}<D_{i+1,j}$).
Fig.~\ref{fig3}D shows the column-wise minima of the distance matrix for the smoothed distance matrix of the noisy simulation of the Lorenz-system, for example.
We use the same kernel smoothing as in the extended thresholded meta RP where the band width is $10$ time steps, half of the first remarkable minimum of $p_{\tau}$ (Fig.~\ref{fig3}B).  
Now horizontal bands are visible containing curves which remember time series of an oscillation with changing amplitude (Fig.~\ref{fig3}D).
At some time steps these curves are connected by diagonal lines which actually build the original RP in Fig.~\ref{fig2}A.

\begin{figure}
\resizebox{0.75\paperwidth}{!}{\includegraphics{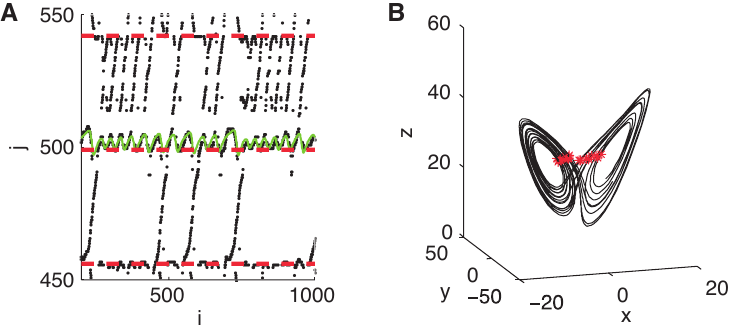}}
\caption{(A) Part of Fig.\ref{fig3}D where the black dots mark the positions of the column-wise local minima. For comparison, the rescaled third component of the undisturbed Lorenz-system is displayed by the green solid line. The red dashed lines underline horizontal levels with an accumulation of black dots. (B) The trajectory of the undisturbed Lorenz-system in its state space (black line). The red stars represent the time stamps which correspond to red dashed lines in A.}
\label{fig4}
\end{figure}

For the Lorenz-example, a selected "band" (at j=500) is presented in Fig.~\ref{fig4}A. 
This horizontal structure coincides well with the z-component of the Lorenz system (App.~\ref{lorenz}).
These bands of black points represent the stay of the trajectory in one of the two wings of the attractor for one period.
The curves are projections of the stay in the other wing, whereas the diagonal lines crossing the empty space display the oscillations in the corresponding wing.
Further, one bound of these bands, characterized by the highest number of black dots, visualizes the closest point of the corresponding wing-internal oscillation to the oscillations in the other wing which is indicated by its location along the merging zone of the trajectory (c.f. Fig.\ref{fig4}B).

\begin{figure}
\resizebox{0.75\paperwidth}{!}{\includegraphics{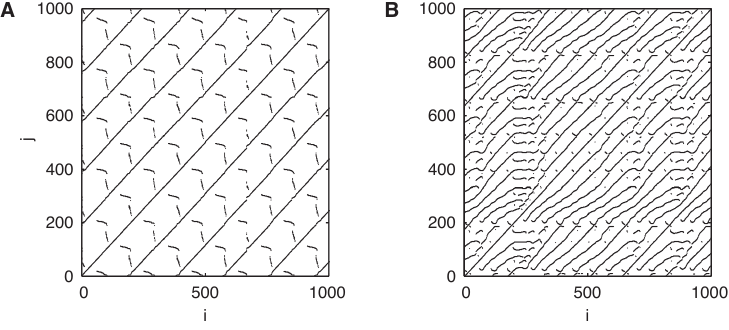}}
\caption{Positions of the column-wise local minima of distance matrices for simulations of: (A) the van der Pol-oscillator (App.~\ref{lorenz}); and (B) the R\"{o}ssler-system.}
\label{fig20}
\end{figure}
So, we only find this horizontal band structure in systems where the trajectory stays in at least two separate regions of the state space for more than one main oscillation.
This is indicated by the counterexamples, the van der Pol oscillator and the R\"{o}ssler system (c.f. Fig.~\ref{fig20}), where the bands disappear or partly vanish, respectively.
The Lorenz system intrinsically fulfills the condition of separate regions in the phase space with separated oscillations, i.e. the trajectory stays in separated state space regions for more than one oscillation, but also multi-stable systems comprise this property.
So, this technique of visualization does not only display a projection of the oscillation in the different parts but also indicates the switching of the trajectory from one part to another.


\section{Application}

We apply the proposed technique of the local minima of the distance matrix to spatio-temporal data of an ecological system, and demonstrate the ability of this approach to visualize driving forces of the system.

\subsection{Data}

The first data set is the estimation of the chlorophyll concentration (CHL, $[$min,max$] = [0.0107,62.3735]$ $mg/m^3$) in the CCS from 1998 to 2016, i.e. 1431 time steps.
It results from the merging of multiple satellite sensor outputs of the ocean color and in situ measurements of CHL in this region \cite{kahru2012},\cite{kahru2015}. 
The spatial resolution of the gridded data is 4 km where the upper-left and the lower-right corners of the grid are given by (45$^{\circ}$N, -140$^{\circ}$E) and (30.03597$^{\circ}$N, -115.5454$^{\circ}$E), respectively, resulting in 61 $\times$ 87 data points.
The temporal resolution is 5 days after averaging.
Missing data are exchanged by a two times linear interpolation which is based on the previous and the following time stamp.
The second data set is the satellite based SST ($[$min,max$] = [10.5,24.45]$ $^{\circ}C$) in the CCS from 1981 to 2016, i.e. 2552 time steps, with the same spatial and temporal resolution as in the CHL data.
In this study, we focus on the Southern Californian Bight (SCB) (upper-left corner: (34.4245$^{\circ}$N, -120.9898$^{\circ}$E); lower-right corner: (32.2662$^{\circ}$N, -117.0880$^{\circ}$E)) schematically drawn in Fig.~\ref{fig0}A.

\subsection{Preprocessing}

\begin{figure}
\resizebox{0.75\paperwidth}{!}{\includegraphics{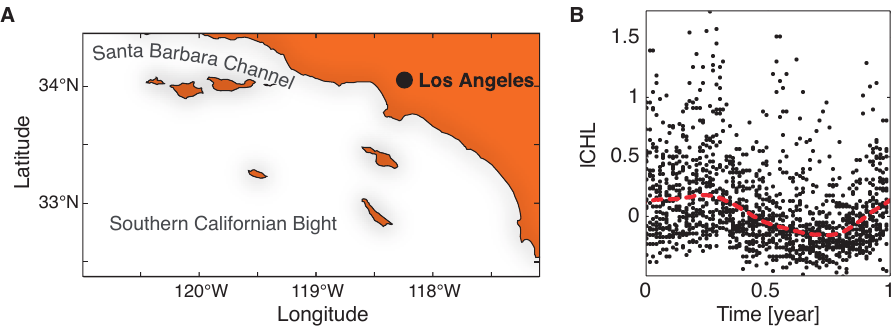}}
\caption{(A) Schematic draw of the considered geographic region, the Southern California Bight, with the location of Los Angeles (LA) and the Santa Barbara Channel (SBC) . The black curves mark coastlines. (B) Exemplary centralized logarithmic chlorophyll concentration  (lCHL), for the pixel at the location (33.2734$^{\circ}$N,-117.7232$^{\circ}$E). The bold line indicates the estimated annual cycle.}
\label{fig0}
\end{figure}

The first step of the preprocessing is the transformation of CHL by means of the decadic logarithm in order to normalize the data.
Next, the logarithmic CHL as well as the SST data are centralized by subtracting the respective grid point-wise temporal median.
This is done in order to reduce local differences in the mean level focusing on the dynamics of both variables.
For example, the shallow water along the coast line and at submarine plateaus in the canyon structure of the continental shelf has remarkable higher values in the CHL because of an increased reflectance in the used spectral band by means of suspended sediments or land and bottom reflections.
Finally, we remove the annual cycle from both data sets.
This is performed for each grid point, separately:(i) wrapping the time axis; (ii) non-parametrically estimation of the annual cycle by means of a robust Loess-regression with a bandwidth which is determined by 50$\%$ of the data points, the quartile range \cite{haerdle1990}; and (iii) calculating the residuals of the fit  (Fig.~\ref{fig0}B).
The residuals of these nonparametric annual models are called anomalies of CHL (CHLA) and anomalies of SST (SSTA).

\subsection{Mapogram-based recurrence plot}

The data of CHLA and SSTA consist of time series of images, i.e. two dimensional data fields with 61$\times$87 pixels.
In order to analyze these given spatio-temporal data by means of the RP framework, we use the mapogram-based recurrence plot (MRP) which is able to provide additional information on the observed system in comparison to a classical RP analysis of spatially averaged data \cite{riedl2015}. 
The MRP is based on the similarity measure $S^m_{t,t'}$ (c.f. Eq.~\ref{eq6}), quantifying the spatial similarity of two images at time stamps $t$ and $t'$ .
As the measure of similarity, we choose the weighted Bhattacharyya distance which is premised on the mapogram (c.f. App.~\ref{similarity}). 
This mapogram is a representation of the image and is determined by two parameters: (i) the binning of the gray-scale values of the image which controls the level of simplification of the spatial pattern; and (ii) the blurring which determines the minimal resolved spatial scale.
Hence, the tuning of the spatial scales of interest allows a multiscale investigation of the image \cite{riedl2015}.
The resulting similarity is used to construct the recurrence matrix $R_{t,t'}$ by means of  
\begin{equation}\label{eq1}
R_{t,t'}(\epsilon)=\Theta(S_{t,t'}-\epsilon).
\end{equation}
$\Theta$ denotes the Heaviside function, 
$\epsilon$ is the threshold of similarity. 
Settings of our analysis are:
(i) $\epsilon$ is determined by a fixed recurrence rate of 5$\%$;
(ii) there are 4 bins where the thresholds between the bins are given by the lower quartile, the median, and the upper quartile of the data; and
(iii) the blurring coefficients of the mapogram were $\{0, 2, 5, 10, 15, 20, 25, 30, 35, 40\}$ pixels, which relate to the minimal resolved spatial scales of about $\{1, 4, 10, 20, 30, 40, 50, 60, 70, 80\}$ pixels, respectively.  
All in all, we have 10 recurrence plots for each data set.
The underlying similarity matrix will be analysed as the distance matrix by means of the proposed approach (Sec.~\ref{newapproach}) since the distance is used to encode the similarity of two states, too, by their closeness in the state space.
The only difference in this application is the use of the local maxima instead of the local minina.


\subsection{Results}

\begin{figure*}
\centering
\resizebox{0.75\paperwidth}{!}{\includegraphics{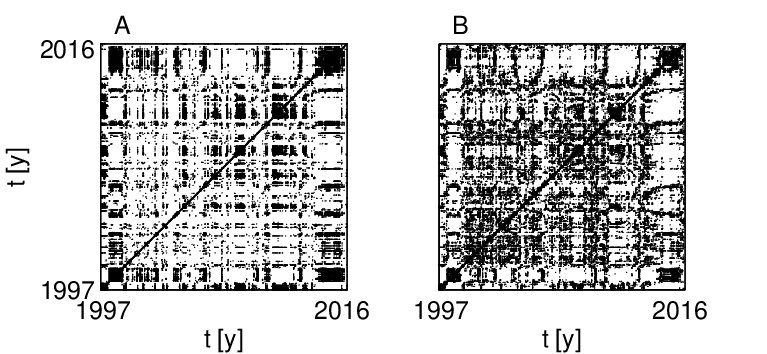}}
\caption{Two mapogram based recurrence plots of the anomaly data of chlorophyl concentration.
They correspond to the smallest resolved spatial scales of 1 pixel (A) and 20 pixels (B).}
\label{fig4a}
\end{figure*}

The recurrence structures within the MRPs of different scales differ significantly (Fig.~\ref{fig4a}), where the largest differences are between scale 1 (4 km) and 20 (80 km).
(Fig.~\ref{fig4a}A, B, respectively).
A further increase of the spatial scale does not lead to qualitative changes.
We further focus on the latter MRP and its underlying similarity matrix taking into account spatial pattern with scales greater than 80km. 

\begin{figure*}
\centering
\resizebox{0.75\paperwidth}{!}{\includegraphics{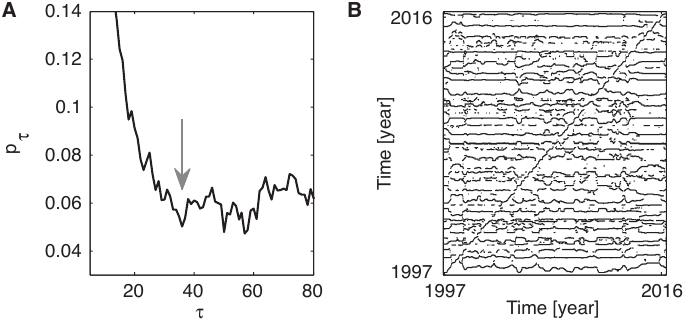}}
\caption{(A) Probability of recurrence after $\tau$ time steps $p_{\tau}$, (Eq.~\ref{eq10}) based on the MRP in Fig.~\ref{fig4a}B. The arrow marks the local minimum of $p_{\tau}$ at 36 which is used to determine the kernel's bandwidth of 18 time steps. (B) Positions of the column-wise local maxima of the smoothed similarity matrix which is the basis of the MRP in Fig.~\ref{fig4a}B. }
\label{fig5}
\end{figure*}

Based on the probability of recurrence after $\tau$ time steps $p_{\tau}$ (Eq.~\ref{eq10}) derived from the MRP in Fig.~\ref{fig4a}B, we select a kernel's bandwidth of 18 time steps for temporal smoothing (derived from the minimum at $\tau = 36$, Fig. 7A). 
This bandwidth corresponds to the smallest resolved temporal scale of 90 days or a quarter of a year.
The underlying similarity matrix of the MRP (Fig.~\ref{fig4a}B) is transformed to the meta-RP  by means of the introduced Epanechnikov-Kernel (Eq.~\ref{kernel}).
After that, the column wise local maxima are selected which are given by values greater than their upper and lower neighbors.
The positions of these local maxima are shown in Fig.~\ref{fig5}B. 

\begin{figure*}
\centering
\resizebox{0.75\paperwidth}{!}{\includegraphics{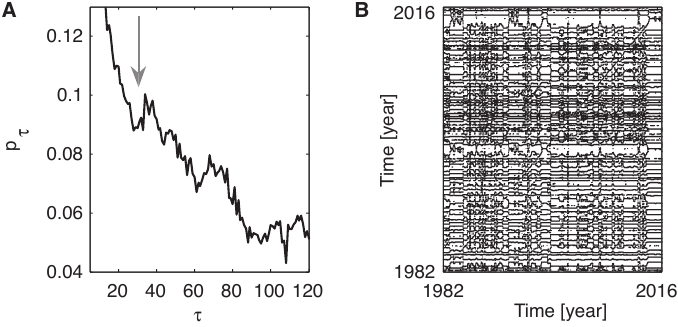}}
\caption{(A) Probability of recurrence after $\tau$ time steps $p_{\tau}$, (Eq.~\ref{eq10}) based on the MRP of the sea surface temperature anomaly data (for spatial scales greater than 20 pixels). The arrow marks the local minimum of $p_{\tau}$ at 29 which is used to determine the kernel's bandwidth of 15 for smoothing the similarity matrix. (B) Positions of the column-wise local maxima of the smoothed similarity matrix of the sea surface temperature anomaly data SSTA.}
\label{fig6}
\end{figure*}

The procedure is then repeated for SSTA.
For comparison, we look at the MRP with the same blurring of 20 Pixel (80km) as in the case of CHLA.
Here, the bandwidth of the kernel for temporal smoothing is 15 time steps corresponding to 75 days (Fig.~\ref{fig6}A).
The positions of the column-wise local maxima of the smoothed similarity matrix are shown in Fig.~\ref{fig6}B.

\begin{figure*}
\centering
\resizebox{0.75\paperwidth}{!}{\includegraphics{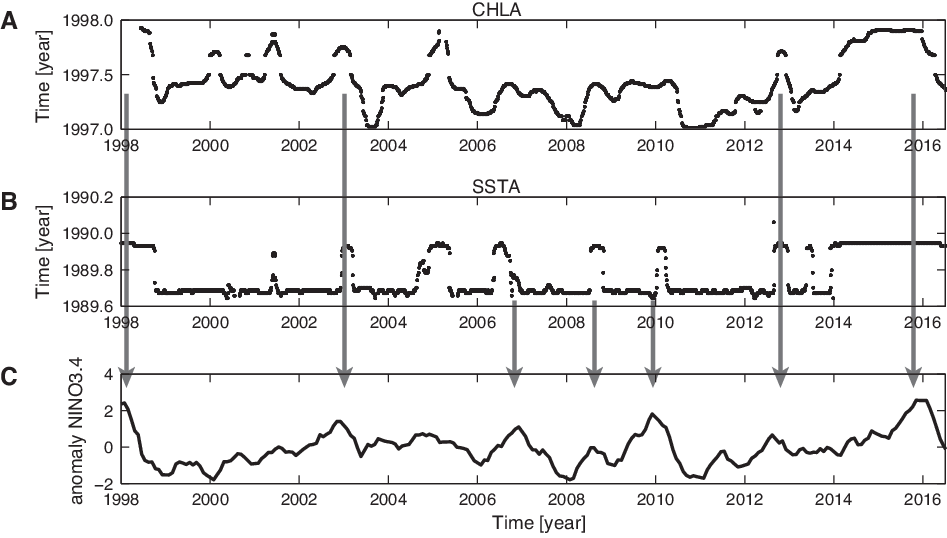}}
\caption{(A) Selected band from Fig.~\ref{fig5}B visualizes the effect of a driving force of $CHLA$. (B) Selected band from Fig.~\ref{fig6}B visualizes the effect of a driving force of $SSTA$. (C) The temporal evolution of the anomaly of NINO3.4, an index of the El Ni\~no-Southern-Oscillation. The black arrows mark simultaneous local maxima in the different time series.}
\label{fig7}
\end{figure*}

In both 
, the local maxima form horizontal band structures (Figs.~\ref{fig5}B,\ref{fig6}B).
In the case of CHLA (Fig.~\ref{fig7}A), we find two global maxima in 1998 and from 2014 to 2016.
In between, the baseline swings down until 2008 or 2009 and rises after.
This baseline is overlaid by several peaks. 
In the case of SSTA (Fig.~\ref{fig7}B), the baseline is rather a rectangular curve with higher values in the temporal range of the global maxima of CHLA (Fig.~\ref{fig7}A).
There are also peaks in between these higher areas with values comparable to the elevated level.
Almost all of these peaks coincide with peaks in CHLA (Fig.~\ref{fig7}A). 
Next we compare these variations with the temporal evolution of the anomaly of the NINO3.4 index, representing the ENSO activity \cite{trenberth1997}.
The enhanced ENSO activity in 2002-2003, 2006, 2008, 2009-2010, and 2012-2013 coincides with the identified maxima in CHLA and SSTA. 
Further, the two global maxima are in the range of the global maxima of CHLA (Fig.~\ref{fig7}A) or the high levels of SSTA (Fig.~\ref{fig7}B).  
Finally, the baseline in between these two global maxima behaves similar to those identified for CHLA (Fig.~\ref{fig7}A) except for the period 1999 to 2002.

\begin{figure*}
\centering
\resizebox{0.75\paperwidth}{!}{\includegraphics{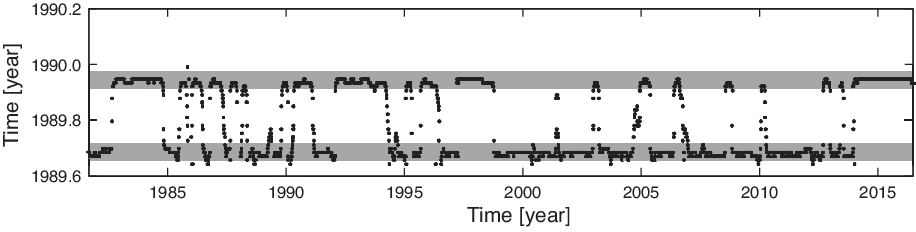}}
\caption{A selected band from Fig.~\ref{fig7}B over the whole sampled period. The gray horizontal bars mark the two different levels of the time series.}
\label{fig8}
\end{figure*}

Finally we consider the whole sampled period of SSTA in order to further investigate the rectangular shape (Fig.~\ref{fig8}).
In the extended temporal range, the rectangular character of the curve continues indicating two regimes of the hydrological system: i) an El Ni\~no like regime occurring from 1997 to 1999, from 1983 to 1985, from 1992 to 1995, and from 2014 to 2016; ii) an La Ni\~na like regime occurring from 1985 to 1990, from 1995 to 1997, and from 1999 to 2014.


\section{Discussion}

In this work, we use the RP approach in an advanced way in order to get a new qualitative view on the trajectory of a spatially extended system which is described by spatio-temporal data.
This new view is given by projections of the trajectories revealing systemic changes which result from driving forces.
It results from the novel extension of the thresholded meta RP reconstructing the driving forces.
We improve this reconstruction by using kernel estimations instead of histograms and focusing on the underlying distance matrix instead of the RP.
So, we are able not only to show the influence of El Ni\~no events on the phytoplankton's growth in the Southern Californian Bight but also 
the documented regime shift in 1998 and the discussed regime shift from 2014 to 2015 by means of the abrupt change between two levels of the constructed curves (Fig.~\ref{fig7},\ref{fig8}).

In particular, the analysis of CHLA and SSTA reveals new hints about long-term effects on the phytoplankton's growth in the Southern California Bight.
So, the coexisting peaks in the projections of CHLA and SSTA and in the ENSO index (Fig.~\ref{fig7}) suggest that the El Ni\~no events, documented for 2002$-$2003,  2009$-$2010 and 2014$-$2016, effect this growth \cite{venrick2012}.
The sea surface temperature seems to be a transmitting link.
This is indicated by the peak in 2005 where no El Ni\~no event was observed but an unusual ocean weather event in the northern California Current leads to responses of the ecosystem similar to that ones of major El Ni\~no events \cite{peterson2006}.
Beside the extremes, the curves in Fig.~\ref{fig7}A and C show similar long-term behavior, except the period from 1999 to 2002, which has not been shown before.
The difference in the period from 1999 to 2002 underlines previous results that strong La Ni\~na, observed from 1998 to 2000 and shown by local minima in Fig.~\ref{fig7}C, do hardly effect the chlorophyll concentration (Fig.~\ref{fig7}A). 
Only the local minimum in 2003 which breaks through the baseline indicates a La Ni\~na like event (Fig.~\ref{fig7}A).
But this event results from a large anomalous intrusion of subarctic water into the CCS \cite{venrick2003}.
A new result is the indication of sharp transitions in the SSTA (Fig.~\ref{fig7}A and Fig.~\ref{fig8}).
That is, the system seems to switch between two levels where the higher one corresponds to El Ni\~no like conditions which is indicated by single peaks coinciding with the events (c.f. Fig.~\ref{fig7}B and C).
We assume that the lower level represents La Ni\~na like conditions.
Three of the four largest periods of the upper level (c.f. Fig.~\ref{fig8}) coincide well with the strongest El Ni\~no events, 1982$-$1983, 1997$-$1998 and 2014$-$2016, during the recorded time \cite{venrick2015}.
But the most impressive part is the very large period of the lower level from 1999 to 2014 which highlights the changes in 1998 and 2014. 
The step from the higher level to the lower one in 1998 (c.f. Fig.~\ref{fig8}) corresponds to a regime shift in the north Pacific ocean \cite{overland2008}.
So we assume that the step in 2014 is a regime shifts, too.

The new findings for this example demonstrate the potential of the proposed method.
The vital condition of this approach is the existence of at least two parts of the state space where the trajectory stays longer than the period of the main cycle.
This condition is given for the shown example of spatio-temporal as well as a large number of other complex systems, e.g. the shown Lorenz-system.   
Therefore, the proposed method promises new insights into other complex systems, too.   


\begin{acknowledgement}
This work was supported by the Volkswagen Foundation (Grant No. 88462), the DFG RTG 2043/1 "Natural Hazards and Risks in a Changing World".
\end{acknowledgement}


\appendix

\section{\label{similarity} Mapogram-based similarity}

The mapogram \cite{nilsson2008} is a representation of the pattern in a gray-scale image and is used to quantify the similarity of two images.
Formally, an image is a two dimensional data field $F=\{f_{ij}\}_{i=1,\dots,N_i;j=1,\dots,N_j}$ where $i$ and $j$ are the indexes of the pixels  and $f_{ij}$ the assigned gray-scale values. 
$N=N_iN_j$ is the total number of pixels. 
The first step constructing the mapogram is a simplification of the gray-scale by means of a histogram
\begin{equation}\label{eq1a}
n_b=\sum_{i=1}^{N_i}{}\sum_{j=1}^{N_j}{g_b(f_{ij})}\end{equation}
where the binary matrices $g_b$ resulted from
\begin{equation}\label{eq2}
g_b(f_{ij})=\left\{ \begin{array}{cc}1 & f_{ij}\in b\textrm{-th bin} \\0 & \textrm{otherwise} \end{array}\right.
\end{equation}
$b=1,\dots,B$ is the index of the histogram's bins, $B$ disjoint right side closed intervals which cover a contiguous part of the gray-scale. 
So $n_b$ is the number of elements with values in the $b$-th bin. 
The $N_i\times N_j$ binary matrices are normalized
\begin{equation}\label{eq3}
m_{b,i,j}=\frac{g_b(f_{ij})}{n_b}\end{equation}
and convoluted with a kernel function $K_{\gamma}$, the blurring, which leads to the mapogram. 
This spatial smoothing is done for each $m_{b,i,j}$ and is controlled by the positive definite non-zero parameter $\gamma$, the band width of the kernel given in units of sample points:
\begin{equation}\label{eq4}
m_{b,\gamma,i,j}=\sum_{i'=1}^{N_i}{}\sum_{j'=1}^{N_j}{m_{b,i',j'}K_{\gamma}\left(\frac{\|(i',j')-(i,j)\|}{\gamma}\right)}
\end{equation}
$K_{\gamma}$ is the Epanechnikov kernel (Eq.~\ref{kernel}) with its bandwidth parameter $\gamma$.
So, the set of $m_{b,\gamma,\bullet,\bullet}$ build the mapogram representing the pattern in one gray-scale image. 
The similarity between the two fields is calculated by a weighting of the Bhattacharyya coefficient, here
\begin{equation}\label{eq6}
S^{m}_{f,f'}(\gamma,B)=\frac{\sum_{b=1}^{B}\sqrt{n_b n'_b}}{\sqrt{(\sum_{b}{n_b})(\sum_{b}{n'_b})}}\frac{\sum_{i=1}^{N_i}\sum_{j=1}^{N_j}\sqrt{m_{b,\gamma,i,j}m'_{b,\gamma,i,j}}}{\sqrt{\sum_{ij}{m_{b,\gamma,i,j}\sum_{ij}{m'_{b,\gamma,i,j}}}}}
\end{equation}
where the second factor is the weight. 
The range of the similarity measure is from 0 to 1, respectively fully dissimilar and equal images. 
For $\gamma\rightarrow 0$, the $m_{b,\gamma,i,j}$ (Eq.~\ref{eq4}) tends to $m_{b,i,j}$ (Eq.~\ref{eq3}).
In the limes $\gamma = 0$, the similarity measure is set to
\begin{equation}\label{eq6a}
S^{m}_{f,f'}(0,B)=\frac{\sum_{b=1}^{B}{\sqrt{n_b n'_b}}}{\sqrt{(\sum_{b}n_b)(\sum_{b}n'_b)}}\sum_{i=1}^{N_i}\sum_{j=1}^{N_j}{\sqrt{m_{b,i,j}m'_{b,i,j}}}
\end{equation}
which corresponds to the known $\kappa$-statistics \cite{monserud1992}.




\section{\label{lorenz} Theoretical models}

First, the Lorenz-system is defined by the system of differential equations:
\begin{equation}
\begin{array}{lcl} 
\dot{x}&=&s(y-x)\\
\dot{y}&=&rx-y-xz\\
\dot{z}&=&-bz+xy.
\end{array}
\end{equation}
The used model parameters are $s=10$, $r=28$, and $b=8/3$ related to a chaotic regime.
It is integrated in the temporal range of $[0,100]$ with the initial condition $(x,y,z)=(1,1,1)$.

Second, the R\"{o}ssler-system is given by:
\begin{equation}
\begin{array}{lcl} 
\dot{x}&=&-y-z\\
\dot{y}&=&x+ay\\
\dot{z}&=&b+z(x-c).
\end{array}
\end{equation}
The values of the parameters are: $a=0.432$, $b=2$, and $c=4$.
The initial condition is $(x,y,z)=(0.1,0.1,0.1)$ and the time span is $[0,1000]$.

Finally, the van der Pol-oscillator is defined by:
\begin{equation}
\begin{array}{lcl} 
\dot{x}&=&a(x-x^3/3-y)\\
\dot{y}&=&x/a.
\end{array}
\end{equation}
The model parameter is set to $a=5$, the initial condition is $(x,y)=(0.5,0)$, and the time span is $[0,1000]$.

For numerical integration of all three dynamical systems, we use the Matlab-function ode45 (Dormand–Prince method).
In order to exclude transient behavior, we only consider values above the 2000th time step.



\begin{thebibliography}{}
\bibitem{eckmann1987}J.P.~Eckmann, S.O.~Kamphorst, D.~Ruelle, Recurrence plots of dynamical systems, Europhys. Lett., \textbf{5}, (1987) 973–977.
\bibitem{marwan2007}N.~Marwan, M.C.~Romano, M.~Thiel, J.~Kurths,Physics Reports, \textbf{438}, (2007) 237–329.
\bibitem{marwan2015}N.~Marwan, J.~Kurth, S.Foerster, Physics Letters A, \textbf{379}, (2015) 894-900.
\bibitem{riedl2015}M.~Riedl, N.~Marwan, J.~Kurths, Chaos: An Interdisciplinary Journal of Nonlinear Science, \textbf{25}, (2015) 123111.
\bibitem{donner2011}R.V.~Donner, M.~Small, J.F.~Donges, N.~Marwan, Y.~Zou, R.~Xiang, J.~Kurths, International Journal of Bifurcation and Chaos, \textbf{21}, (2011) 1019-1046.
\bibitem{casdagli1997}M.C.~Casdagli, Physica D, \textbf{108}, (1997) 12-44.
\bibitem{wang2015}D.~Wang, T.C.~Gouhier, B.A.~Menge, A.R.Ganguly, Nature, \textbf{518}, (2015) 390-394.
\bibitem{venrick2012}E.L.~Venrick, Progress in Oceanography, \textbf{104}, (2012) 46-58.
\bibitem{kahru2012}M.~Kahru, R.M.~Kudela, M.~Manzano-Sarabia, B.G.~Mitchell, Deep Sea Research Part II: Topical Studies in Oceanography, \textbf{77}, (2012) 89-98.
\bibitem{kahru2015}M.~Kahru, R.M.~Kudela, C.R.~Anderson, B.G.~Mitchell, IEEE Geoscience and Remote Sensing Letters, \textbf{12}, (2015) 2282-2285.
\bibitem{haerdle1990}W.~H\"ardle, \textit{Applied Nonparametric Regression (No. 19)} (Cambridge University Press 1990).
\bibitem{silverman1986}B.W.~Silverman, \textit{Density Estimation for Statistics and Data Analysis} (Chapman \& Hall/CRC, London 1986) 42-43.
\bibitem{peterson2006}B.~Peterson et al., State Of The California Current CalCOFI Rep., \textbf{47}, 2006.
\bibitem{venrick2003}E.~Venrick et al., The California Current, \textbf{44}, (2003).
\bibitem{venrick2015}E.~Venrick, CalCOFI Rep., \textbf{56}, (2015).
\bibitem{overland2008}J.~Overland, S.~Rodionov, S.~Minobe, N.~Bond, Progress in Oceanography, \textbf{77}, (2008) 92–102.
\bibitem{monserud1992}R.A.~Monserud, R.~Leemans, Ecol. Modell., \textbf{62}, (1992) 275.
\bibitem{nilsson2008}M.~Nilsson, J.S.~Bartunek, J.~Nordberg, I.~Claesson, ICIP , \textbf{973}, (2008).
\bibitem{trenberth1997}K.E.~Trenberth, Bulletin of the American Meteorological Society, \textbf{78}, (1997).
\end{thebibliography}
\end{document}